\documentclass[journal=nalefd,manuscript=letter]{achemso}
\usepackage[T1]{fontenc}
\usepackage{amsmath,amssymb}
\usepackage{siunitx}
\usepackage[normalem]{ulem}

\DeclareSIUnit{\belmilliwatt}{Bm}
\DeclareSIUnit{\dBm}{\deci\belmilliwatt}
\DeclareSIUnit{\torr}{Torr}

\author{Henry C. Hammer}
\affiliation[The University of Iowa]{Department of Physics \& Astronomy, The University of Iowa, Iowa City, IA 52242}

\author{Hassan A. Bukhari}
\affiliation[The University of Iowa]{Department of Physics \& Astronomy, The University of Iowa, Iowa City, IA 52242}

\author{Yogendra Limbu}
\affiliation[The University of Iowa]{Department of Physics \& Astronomy, The University of Iowa, Iowa City, IA 52242}

\author{Brett M. Wasick}
\affiliation[The University of Iowa]{Department of Physics \& Astronomy, The University of Iowa, Iowa City, IA 52242}

\author{Christopher Rouleau}
\affiliation[ORNL]
{Center for Nanophase Materials Sciences, Oak Ridge National Laboratory, Oak Ridge, Tennessee 37831, USA}

\author{Michael E. Flatt\'e}
\affiliation[The University of Iowa]{Department of Physics \& Astronomy, The University of Iowa, Iowa City, IA 52242}
\alsoaffiliation[TU Eindhoven]{Department of Applied Physics, Eindhoven University of Technology, Eindhoven, The Netherlands}

\author{Durga Paudyal}
\affiliation[The University of Iowa]{Department of Physics \& Astronomy, The University of Iowa, Iowa City, IA 52242}
\email{durga-paudyal@uiowa.edu}

\author{Denis R. Candido}
\affiliation[The University of Iowa]{Department of Physics \& Astronomy, The University of Iowa, Iowa City, IA 52242}
\email{denis-candido@uiowa.edu}

\author{Ravitej Uppu}
\affiliation[The University of Iowa]{Department of Physics \& Astronomy, The University of Iowa, Iowa City, IA 52242}
\email{ravitej-uppu@uiowa.edu}
\title[]{Quantum Coherence of Rare-Earth Ions in Heterogeneous Photonic Interfaces}
\abbreviations{}

\begin{document}

\begin{abstract}
\noindent Harnessing rare-earth ions in oxides for quantum networks requires integration with bright emitters in III-V semiconductors, but local disorder and interfacial noise limit their optical coherence.
Here, we investigate the microscopic origins of the ensemble spectrum in Er$^{3+}$:TiO$_2$ epitaxial thin films on GaAs and GaSb substrates.
\textit{Ab initio} calculations combined with noise-Hamiltonian modeling and Monte Carlo simulations quantify the effects of interfacial and bulk spin noise and local strain on erbium crystal-field energies and inhomogeneous linewidths.
Photoluminescence excitation spectroscopy reveals that Er$^{3+}$ ions positioned at increasing distances from the III-V/oxide interface produce a systematic blue shift of the $Y_1\rightarrow Z_1$ transition, consistent with strain relaxation predicted by theory.
Thermal annealing produces a compensating redshift and linewidth narrowing, isolating the roles of oxygen-vacancy and gallium-diffusion noise.
These results provide microscopic insight into disorder-driven decoherence, offering pathways for precise control of hybrid quantum systems for scalable quantum technologies.
\end{abstract}


%%%%%%%%%%%%%%%%%%%%% Introduction %%%%%%%%%%%%%%%%%%%%%%%%%%%

%\section{Introduction}
Quantum network architectures increasingly rely on solid-state quantum memories coupled to photon emitters. \cite{Wehner2018, Azuma2023, Atature2018, Awschalom2018}
Rare-earth ions (REIs) in crystalline oxides offer exceptional optical and spin coherence because their intra-4\textit{f} transitions are well shielded from the host lattice. \cite{Thiel2011, Zhong2019, Goldner2025}
In particular, Er$^{3+}$ provides telecom-band transitions compatible with existing fiber networks, while III-V semiconductors supply bright, deterministic photon sources and scalable nanophotonic platforms.\cite{Uppu2020, Tomm2021}
Integrating these material classes enables hybrid quantum nodes \cite{strocka2025} that combine scalable photonic circuitry with coherent spin-photon interfaces.\cite{Uppu2021, Yu2023a}
Recent demonstrations of epitaxial Er$^{3+}$:TiO$_2$ on Si\cite{Dibos2022, Ji2024, Singh2024, pettit2025} and III-V\cite{Hammer2025} substrates have extended previous results in a bulk crystal environment\cite{Phenicie2019}, establishing a promising route toward such hybrid quantum architectures.

Maximizing optical coupling between REIs and photonic modes in hybrid photonic integrated circuits requires positioning REIs within tens of nanometers of the heterovalent oxide/III-V interface.\cite{Wu2023, Uysal2025}
Its interfacial lattice and valence mismatches create strain and promote cation interdiffusion, leading to local electric- and magnetic-field fluctuations and ultimately limiting REI coherence. \cite{Wu2023, Ourari2023}
Understanding the microscopic origin and spatial dependence of such noise is, therefore, essential for optimizing hybrid quantum systems.
While decoherence mechanisms in bulk or implanted crystals have been studied extensively \cite{Bottger2006, Sun2008, Bottger2009}, there is a critical need to understand interfaces of heteroepitaxial films and the interplay between interfacial and bulk noise sources.

Here, we develop a microscopic theoretical-experimental framework linking interfacial chemistry to optical decoherence in Er$^{3+}$:TiO$_2$ thin films grown on Ga-based substrates. 
By combining photoluminescence-excitation spectroscopy of $\delta$-doped films with noise modeling and \textit{ab initio} crystal-field calculations, we quantify how gallium diffusion, oxygen-vacancy gradients, and strain collectively shape the spectral response of ensembles of Er$^{3+}$ ions. 
This unified description connects local and ensemble linewidth physics within a single parameter set and provides quantitative guidelines for designing coherent oxide/semiconductor heterostructures that host rare-earth ions.

\begin{figure}[ht!]
\centering
\includegraphics[width=\textwidth]{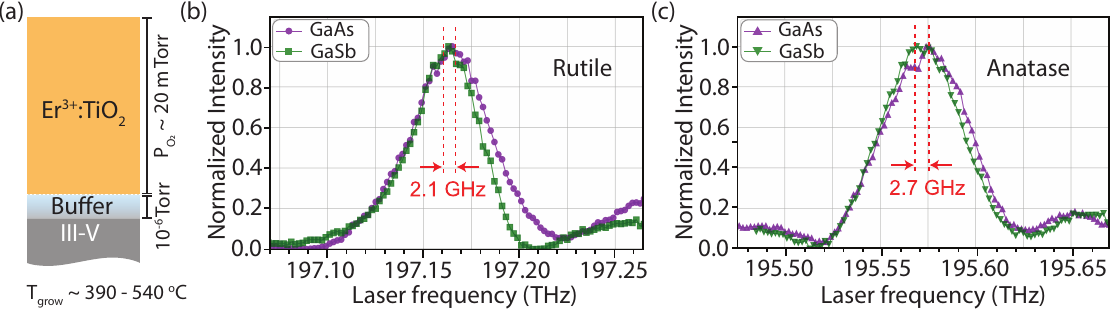}
\caption{(a) Schematic of the synthesized heteroepitaxial Er$^{3+}$:TiO$_2$ thin film on III-V substrates featuring an undoped buffer layer grown under high vacuum. Photoluminescence excitation (PLE) spectra of the $Y_1\rightarrow Z_1$ transition in Er$^{3+}$ doped into the rutile (b) and anatase (c) phase thin films grown on GaAs and GaSb substrates.}
\label{fig:fig1}
\end{figure}

Er$^{3+}$-doped TiO$_2$ thin films were grown on GaAs and GaSb substrates by pulsed laser deposition. 
TiO$_2$ provides a model system for probing REI optical coherence owing to its wide bandgap transparency, minimal host nuclear spin density, and established low-temperature epitaxy on GaAs and GaSb \cite{Hammer2025}.
Figure~\ref{fig:fig1}(a) schematically illustrates the heterogeneous film geometries featuring a thin, undoped TiO$_2$ buffer ($\approx 1-5~\si{\nano\meter}$) grown under high vacuum and a 3000 ppm Er-doped film of $60-90~\si{\nano\meter}$ thickness grown in an oxygen ambient pressure of 20~\si{\milli\torr}.
Growth temperature and surface-preparation conditions determine whether rutile (R) or anatase (A) polymorphs form.
R-TiO$_2$ films grown on GaAs(100) develop mixed (110)/(210) orientations that partially relax strain, whereas growth on GaSb(100) yields a (200) orientation with an estimated interfacial compressive strain of 0.48\%, based on minimal coincident interface area calculations.
In contrast, A-TiO$_2$ films grow with (001) orientation on both GaAs(100) and GaSb(100) substrates, exhibiting biaxial in-plane compressive strains of 0.43\% and 0.12\%, respectively.\cite{Hammer2025} 
The sensitivity of TiO$_2$ polymorphs to substrate lattice mismatch thus enables strain engineering through substrate selection, offering a controllable handle on interfacial stress and defect formation.
These distinct strain states make the A-TiO$_2$/III-V system an ideal platform for probing how lattice mismatch and interface chemistry influence the REI optical spectra.

Elements of the optical response of Er$^{3+}$ doped into TiO$_2$ thin films can be predicted using crystal field calculations.
We compute\cite{limbu2025ab} crystal-field coefficients (CFCs) using large supercells containing a single Er$^{3+}$ substituting a Ti$^{4+}$, corresponding to a 1.389\% doping concentration in both rutile ($3\times3\times4$ supercell) and anatase ($3\times 3 \times 2$ supercell) polymorphs.
The CFCs are extracted\cite{limbu2025ab} from the crystal-field potential obtained using non-spin-polarized density-functional theory (DFT) using the $4f$-core approximation~\cite{novak2013crystal} and dielectric constants of 2.00 (rutile)\cite{limbu2025ab} and 2.231 (anatase)\cite{limbu2025ab,paudyal2025singly}, which are required for determining the $4f$ energy levels. 
These coefficients serve as inputs to an effective Hamiltonian solved using the $\textit{qlanth}$ code, yielding the Er$^{3+}$ $4f$ energy spectrum\cite{qlanth,qlanthpr}.
Epitaxial strain lowers the local Er site symmetry in the ideal lattice from D$_{2h}$ to C$_{2h}$ in rutile and from D$_{2d}$ to C$_{s}$ in anatase on both (GaAs and GaSb) substrates. 
This results in nine non-zero CFC ($B^k_q$) terms, leading to eight $Z_1$ - $Z_8$ ground-state and seven $Y_1$ - $Y_7$ first-excited-state doublets (see Supplemental Material).
DFT computations predict characteristic dependences of the level splitting on the presence of local defects ({\it e.g.,} oxygen vacancies) and strain relaxation, as will be discussed later.

Figures~\ref{fig:fig1}(b,c) show photoluminescence excitation (PLE) spectra measured at 5~\si{\kelvin}, highlighting the characteristic shift in the $Y_1 \rightarrow Z_1$ transition between the two polymorphs: 197.165~\si{\giga\hertz} (1520.5~\si{\nano\meter}) on R-TiO$_2$ and 195.575~\si{\giga\hertz} (1532.9~\si{\nano\meter}) on A-TiO$_2$.
Growth on different substrates further shifts the $Y_1 \rightarrow Z_1$ transition by about 2.5~\si{\giga\hertz}, in good agreement with the predictions from crystal-field calculations (see Supplemental Material).
The R-TiO$_2$ film grown on GaSb exhibits a more pronounced asymmetric lineshape than that on GaAs.
Although A-TiO$_2$ experiences different absolute strain levels than R-TiO$_2$, the fractional change in strain between films grown on GaSb and GaAs is comparable, as reflected by similar frequency shifts.
Despite this, the measured PLE spectra are more symmetric and exhibit a slightly narrower inhomogeneous linewidth on GaSb ($49\pm1~\si{\giga\hertz}$) than on GaAs ($53\pm1~\si{\giga\hertz}$).
Such substrate-dependent asymmetry and linewidth variations indicate that multiple factors, such as strain, intermixing, and vacancies, act concurrently to modify the local environment of Er$^{3+}$ ions near the oxide/III-V interface.
These effects are expected to vary spatially across the film thickness and ultimately influence the measured optical lineshapes.

\begin{figure}[ht!]
\centering
\includegraphics[width=\textwidth]{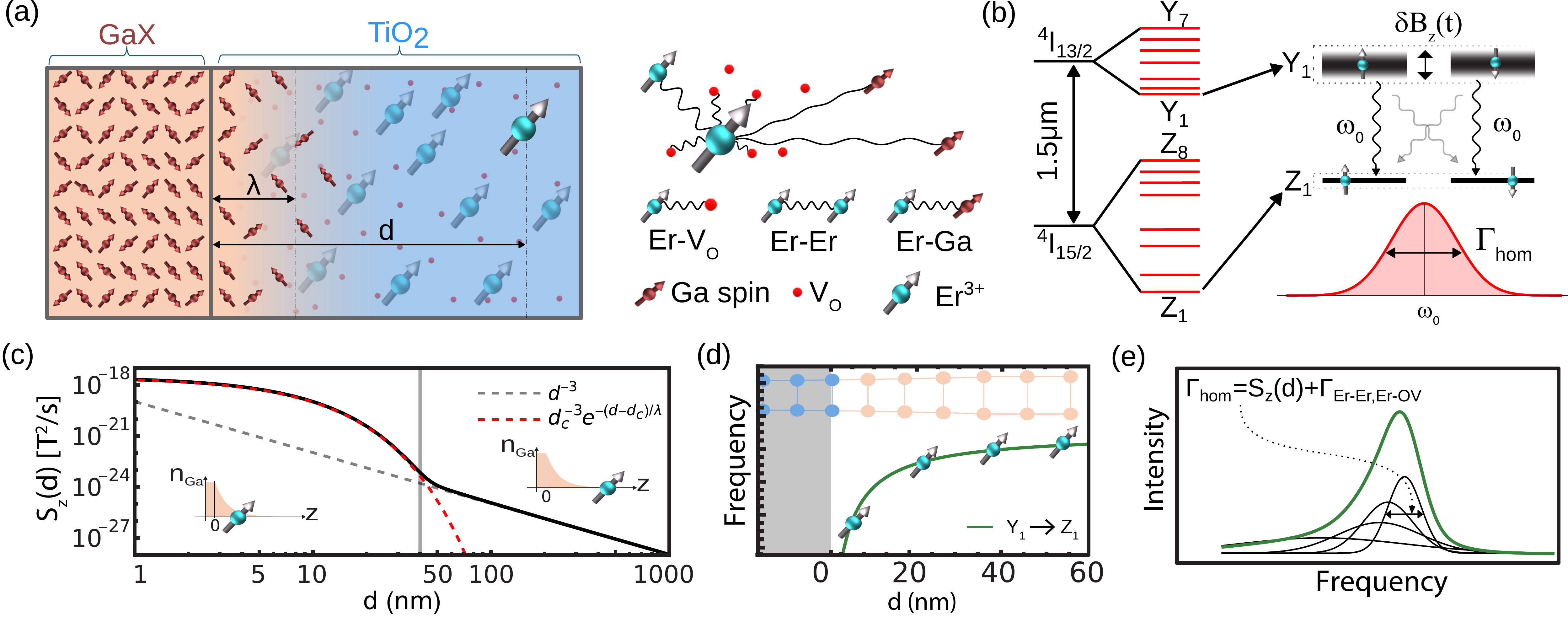}
\caption{(a) Schematic of the GaX-TiO$_{2}$ heterostructure indicating Er ions (cyan) at interfacial distance $d$ that interact with Ga spins (brown) that diffuse into TiO$_2$ with characteristic length $\lambda$ and oxygen vacancies (red) $V_\textrm{O}$. 
(b) Illustration of linewidth broadening arising from the fluctuating magnetic fields $\delta B_z(t)$ on the $Y_{1}\rightarrow Z_{1}$ transition of the crystal-field-split energy levels of Er$^{3+}$.
(c) Spectral noise density $S_z(d)$ due to Ga nuclear spins as a function of distance $d$, compared to the expected $d^{-3}$ (gray dashed) and $d_c^{-3}e^{-(d-d_c)/\lambda}$ (red dotted) dependencies with $\lambda=4$ nm and $d_c=0.35$ nm. Insets show a schematic of Ga spin density distributions near and far from the interface.
(d) Distance-dependent $Y_{1}\rightarrow Z_{1}$ transition frequency $\omega(d)$, illustrating the influence of interfacial strain.
(e) Illustration of the asymmetric inhomogeneous lineshape resulting from the interplay between strain and distance-dependent individual Er$^{3+}$ homogeneous linewidths.
}
\label{fig:fig2}
\end{figure}

Figure~\ref{fig:fig2} summarizes how interfacial intermixing, point defects, and strain jointly influence the Er$^{3+}$ optical transitions and determine the measured inhomogeneous linewidth.
These factors not only broaden the homogeneous linewidth $\Gamma_\textrm{hom}$ of individual ions but also induce static or dynamic frequency shifts that vary with the interfacial distance $d$.
The resulting noise environment includes interactions of Er$^{3+}$ with Ga, oxygen vacancies ($V_\textrm{O}$), and other Er ions, as indicated in Fig.~\ref{fig:fig2}(a).
Interfacial intermixing, wherein gallium atoms from the substrate diffuse into the oxide lattice over a characteristic diffusion length $\lambda$, perturbs the local electrostatic potential while simultaneously introducing both nuclear spins and unpaired electronic spins and associated noise.
Oxygen vacancies are formed predominantly during growth due to limited adatom mobility at the relatively low substrate temperatures ($<400~\si{\celsius}$) and the non-equilibrium plasma plume kinetics. $V_\textrm{O}$ occupation dynamics generate fluctuating electric and magnetic fields that drive charge- and spin-state fluctuations of nearby Er$^{3+}$ ions \cite{Lindera2017, Singh2024, Candido2021b, Candido2024, Grant2024}. 
In addition, the relatively high Er concentration ($\sim$3000~ppm) produces strong dipolar Er--Er interactions that contribute to spectral diffusion and define a baseline inhomogeneous linewidth.\cite{Bottger2006, Singh2024, Grant2024}

Near the heterovalent interface, a significant contribution to the Er$^{3+}$ linewidth arises from the electromagnetic noise generated by Ga atoms.
Within a few nanometers of the interface, the Er$^{3+}$ ions reside in a dense bath of Ga impurities, whose nuclear spins produce slowly fluctuating magnetic fields, leading to substantial broadening of the Er$^{3+}$ transitions.
Although Ga atoms migrating into TiO$_2$ could form paramagnetic centers, such configurations are expected to be extremely rare.
In TiO$_2$, Ga$^{3+}$ predominantly substitutes on the Ti$^{4+}$ site, and charge compensation via nearby oxygen vacancies drives the resulting Ga-V$_\textrm{O}$ complexes towards closed-shell electronic configurations that do not host unpaired electrons \cite{Okajima_2006, Gionco2015}.
Any transient electron-spin centers that may form under non-equilibrium interfacial conditions would possess large dipole moments and therefore couple strongly to their environment, leading to nanosecond or faster spin-fluctuation dynamics \cite{Wu2021}, far shorter than the correlation time of nuclear spins \cite{Hall2014, DeSousa2003}.
Consequently, such a dilute concentration of rapidly fluctuating electron-spin noise results in a negligible contribution to the Er$^{3+}$ linewidth.
For these reasons, and as discussed further in Supplemental Material, we neglect the contribution of Ga-related unpaired electrons in the linewidth analysis.

Far from the interface, the local Ga nuclear-spin bath becomes increasingly dilute, allowing it to be treated as a remote, fluctuating three-dimensional spin bath.
To describe this effect, we model the position-dependent spectral noise density, $S_z (d)$, and the corresponding linewidth broadening $\Gamma_{\textrm{Ga}}(d)=\gamma_{\textrm{Er}}^2 S_z(d)/\pi$ with $\gamma_{\textrm{Er}}$ the gyromagnetic ratio of Er ions, as illustrated in Fig.~\ref{fig:fig2}(b) (see Supplemental Material). 
For distances $d\gg\lambda$, the noise follows the dipolar scaling $S_{z}(d)\propto n_{\textrm{Ga}}/d^3$, where $n_{\textrm{Ga}}$ is the density of the three-dimensional bath of Ga nuclear spins of the substrate\cite{Pham2016, Zhang2017}.
For $d\gtrsim \lambda$, the distance dependence transitions to $S_{z}(d) \propto n_{\textrm{Ga}}e^{-(d-d_c)/\lambda}/d_c^3$, where $d_c$ corresponds to the shortest Ti-Ti separation in TiO$_{2}$ and defines the minimum possible distance between a diffused Ga atom and an Er$^{3+}$ ion (Fig.~\ref{fig:fig2}(c)). 
This exponential attenuation captures the rapid rise of magnetic noise near the interface and its gradual decay into the film interior, deviating strongly from the $1/d^{3}$ behavior expected for purely bulk spin noise (gray dashed line).

In addition to magnetic and defect-related noise, local strain plays a central role in shaping the optical response.
As shown schematically in Figure~\ref{fig:fig2}(d), interfacial lattice mismatch introduces compressive strain that relaxes with increasing distance from the interface, leading to position-dependent Stark shifts of the Er crystal-field levels. 
When position-dependent frequencies and linewidths coexist, the ensemble spectrum can acquire an intrinsic asymmetry.
This interplay, visualized in Figure~\ref{fig:fig2}(e), provides a microscopic origin for the asymmetric lineshapes observed in Figure~\ref{fig:fig1}(b,c) and motivates a spatially resolved investigation of the underlying strain and noise fields.

\begin{figure}[htpb!]
\centering
\includegraphics[width=\textwidth]{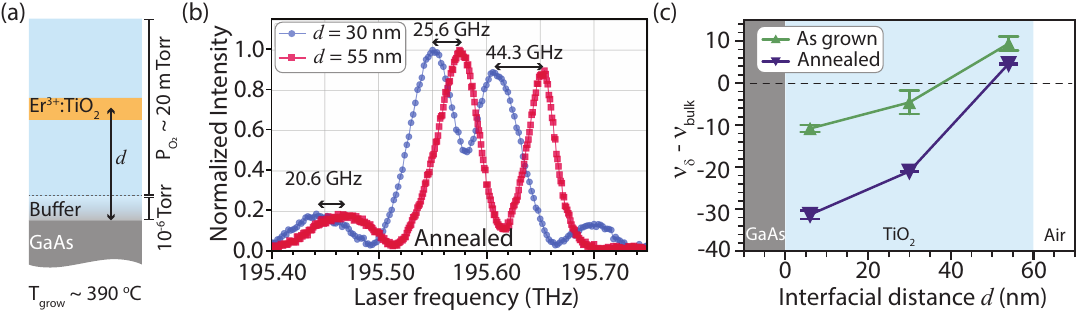}
\caption{(a) Schematic of the selectively-doped Er$^{3+}$:TiO$_2$ thin film on GaAs featuring a $2~\si{\nano\meter}$-thin doped layer introduced at an interfacial distance $d$. (b) PLE spectra around the $Y_1\rightarrow Z_1$ transition of Er$^{3+}$ ions in such annealed thin films exhibit significant frequency shifts due to distance-dependent strain in the thin film. (c) Frequency shift of the $Y_1\rightarrow Z_1$ transition ($\nu_\delta$) before and after annealing, relative to the measured frequency in uniformly-doped samples ($\nu_\textrm{bulk}$), highlights the interplay between strain and oxygen-vacancy contributions.}
\label{fig:fig3}
\end{figure}

We synthesized pseudo-$\delta$-doped TiO$_2$ films containing a narrow (2 nm) Er-doped layer positioned at a controlled distance from the GaAs/TiO$_2$ interface, as illustrated in Figure \ref{fig:fig3}(a).
These $\delta$-doped structures isolate a well-defined spatial slice of the strain and intermixing profiles, enabling direct correlation between local environment and optical response.
We focus on A-TiO$_2$ grown on GaAs because this combination yielded the most reproducible crystalline quality, sub-nm surface roughness, and short Ga diffusion length ($\lambda\approx4~\si{\nano\meter}$) under the low growth temperatures ($<$400\si{\celsius}) required for rare-earth incorporation.\cite{Hammer2025}
While similar epitaxy on GaSb is feasible, the narrower process window and higher volatility of Sb make GaAs a more robust substrate for controlled interfacial chemistry and strain engineering using pulsed laser deposition.
We investigated the optical properties of the $\delta-$doped samples before and after annealing ($4~\si{\hour}$ at $450\si{\celsius}$ in air) to probe the influence of oxygen vacancies~\cite{Grant2024}.

PLE spectra of annealed samples with Er$^{3+}$ doped at two interfacial distances, $d = 30$ and 55~\si{\nano\meter}, are shown in Figure~\ref{fig:fig3}(b).
The spectra feature three prominent peaks, with the central one corresponding to the $Y_1 \rightarrow Z_1$ transition that blueshifts by $25.6~\si{\giga\hertz}$ with increasing $d$.
Furthermore, distinct blueshifts are observed for peaks corresponding to different crystal-field-split transitions, in agreement with \textit{ab initio} computations (see Supplemental Material). 
The distance-dependent frequency shift $\nu_{\delta}(d)$, plotted in Figure \ref{fig:fig3}(c) before and after annealing, follows a quadratic dependence.
After annealing, we observe a steeper relative shift in $\nu_\delta(d)$, potentially signaling increased interfacial strain from additional Ga diffusion.
Additionally, a slight redshift of $5~\si{\giga\hertz}$ is observed between the annealed and as-grown samples far from the interface ($d=55~\si{\nano\meter}$), which potentially arises from the annihilation of $V_\textrm{O}$s as demonstrated via simulations of reaction-diffusion model (see Supplemental Material).

\begin{figure}[htpb!]
\centering
\includegraphics[width=\textwidth]{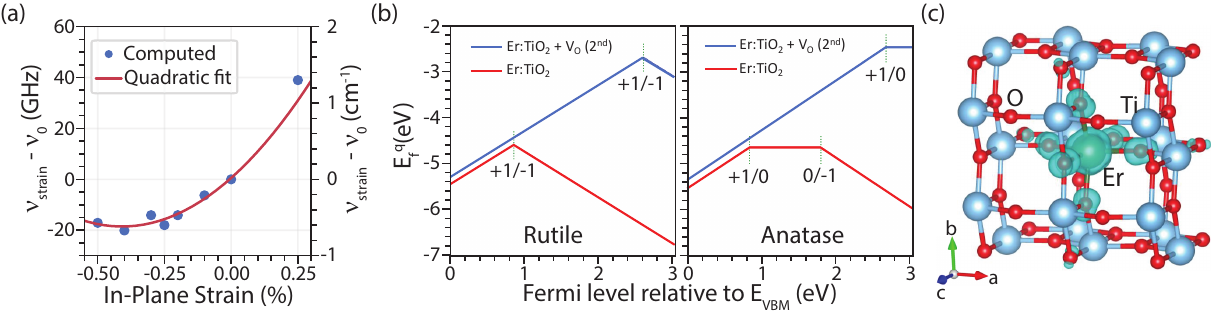}
\caption{(a) Effect of in-plane compressive and tensile strain (with 1\% fixed $c$-axis compression) on crystal field energy levels. (b) Defect formation energy of Er-doped TiO$_2$ with (blue) and without (red) $V_O$ at the next nearest neighbor. (c) Spin density of neutral Er-doped A-TiO$_2$ ($q=0$). The large spin density on Er arises from its $\sim$3~$\mu_B$ spin moment (Er$^{3+}$), while nearby O atoms show small induced spin densities due to Er-4\textit{f} and O-$2p$ hybridization (see Supplemental Material).}
\label{fig:fig4}
\end{figure}

The microscopic origin of the distance-dependent frequency shifts was predicted by DFT for Er-doped rutile and anatase TiO$_2$, both with and without nearby oxygen vacancies.
First, we compute how biaxial strain modifies the crystal-field splitting and transition energies of Er$^{3+}$ in A-TiO$_2$ without $V_\textrm{O}$.
Using the experimentally determined out-of-plane (c-axis) strain of $-1$\% (compressive) from X-ray diffraction measurements, we varied the in-plane biaxial strain uniformly between compressive and tensile values.
Figure~\ref{fig:fig4}(a) reveals that relaxing from an in-plane compressive strain of $-0.45$\% (consistent with epitaxial conditions) to zero strain produces a 20~\si{\giga\hertz} blueshift of the $Y_1 \rightarrow Z_1$ transition, in close agreement with the $\delta$-doped measurements (Figure~\ref{fig:fig3}(c)).
The computed strain-dependent frequency shift in Fig.~\ref{fig:fig4}(a) exhibits a quadratic dependence for small strains ($\pm$0.5\%).
Minor changes in out-of-plane strain after annealing may account for the slightly broader frequency range observed experimentally in Figure~\ref{fig:fig3}(c).
Because the DFT-calculated frequency variation with strain and the measured frequency shift with interfacial distance (Figure~\ref{fig:fig3}(c)) are both quadratic, we conclude that the in-plane strain relaxes linearly with $d$.
This correspondence both reinforces our experimental observation and provides a physically grounded model for distance-dependent frequency shift as: 
\begin{equation}
\omega_\textrm{strain}(d) \sim \omega_0-a d^2.
\end{equation}

Alongside strain, charge-state compensation arising from Er$^{3+}$ substituting for Ti$^{4+}$ and the stability of associated $V_\textrm{O}$ defects govern the local crystal-field environment, influencing both the transition frequencies and linewidth broadening~\cite{Martins2024}.
When a first-shell (nearest-neighbor) oxygen vacancy is introduced in R-TiO$_2$, the Er$^{3+}$ local site symmetry reduces to C$_1$ from C$_{2h}$, resulting in fifteen non-zero $B^k_q$ values and a relatively large $4f$ splitting compared to experiments.
Consequently, second-shell vacancies that result in C$_s$ site symmetry with nine non-zero $B^k_q$ values were considered as more representative of the measured spectra.
The computed crystal-field coefficients ($B^k_q$) and resulting intra-4$f$ level energies for both R-TiO$_2$ and A-TiO$_2$ are discussed in detail in Supplemental Material.
$V_\textrm{O}$ defect-formation-energy calculations (Figure~\ref{fig:fig4}(b)) reveal different charge compensation behavior in rutile and anatase TiO$_2$.
The formation energy of a defect in charge state $q$ was computed as $E_\text{f}^\text{q}(\epsilon_\text{F}) = E_{ds}^q -  E_{ps}^q + \sum_i n_i\mu_i + q(E_\text{VBM} + \epsilon_\text{F}) + E_\text{corr}$, where $E_{ds}^q$ ($E_{ps}^q$) is the total energy of the defective (pristine) supercell, $\mu_i$ the chemical potentials at site $i$, $n_i = 1$ ($-1$) for an added (removed) $i^{th}$ atom, $E_\text{VBM}$ the valence band maxima, $\epsilon_\text{F}$ the Fermi energy, and $E_\textrm{corr}$ the electrostatic and image-charge corrections.\cite{limbu2025stability}
In R-TiO$_2$, isolated Er favors a $q=-1$ charge state at mid-gap, but the presence of a $V_\textrm{O}$ at the second shell stabilizes the $q=+1$ configuration up to a Fermi energy of 2.615 eV.
The neutral state ($q=0$) is unstable, indicating a preference for \textit{local} charge compensation through proximal $V_\textrm{O}$ that enhances local electric- or magnetic-field fluctuations and reduces optical coherence.
In contrast, Er-doped A-TiO$_2$ favors a neutral ($q=0$) charge state, with the $q=+1$ appearing only when a $V_\textrm{O}$ is introduced nearby, highlighting \textit{non-local} charge compensation through the extended oxygen network as seen in Figure~\ref{fig:fig4}(c).
These charge-state differences modulate the crystal-field coefficients and the 4\textit{f}-level splitting (see Supplementary Material), while changes in their occupation lead to charge noise that determines linewidth broadening.

\begin{figure}
\centering
\includegraphics[width=0.5\textwidth]{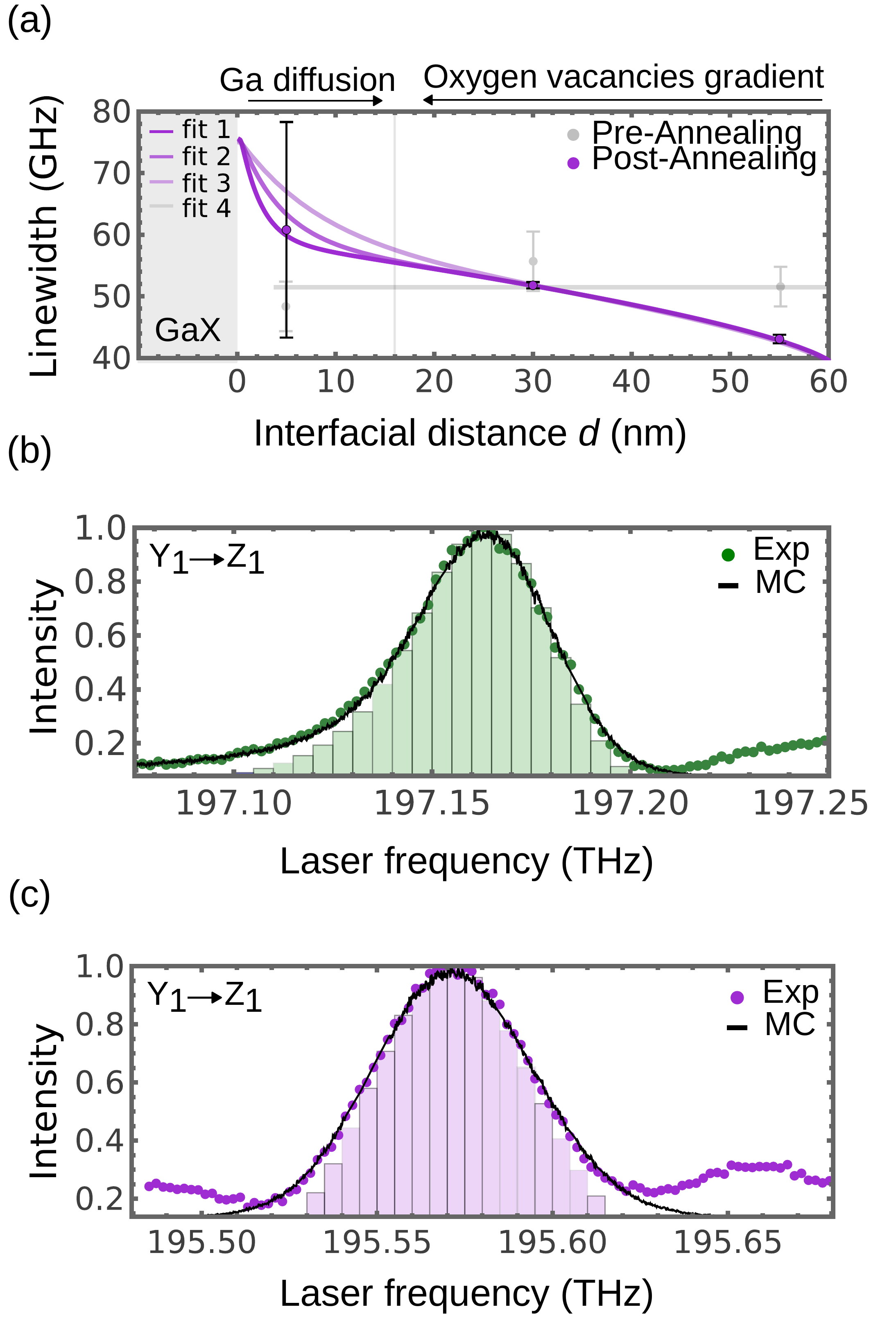}\caption{(a) Inhomogeneous linewidth $\Gamma_{inh}$ as a function of distance from GaX-TiO$_2$ interface before and after annealing of the $\delta$-doped $\mathrm{A\text{-}TiO_2}$ samples. (b) Experimental and simulated lineshape for the $Y_{1}\rightarrow Z_{1}$ transition in $\mathrm{A-TiO_{2}}$. (c) Experimental and Monte Carlo–simulated lineshape for the $Y_{1}\rightarrow Z_{1}$ transition in $\mathrm{R-TiO_{2}}$. }
\label{fig:fig5}
\end{figure}

Having established that Ga intermixing (Figure~\ref{fig:fig2}), strain relaxation, and vacancy-mediated charge compensation jointly define the local environment of Er$^{3+}$, we now quantify their combined impact on the optical linewidths.
Figure~\ref{fig:fig5}(a) presents the distance-dependent inhomogeneous linewidth of the $Y_1 \rightarrow Z_1$ transition obtained from A-TiO$_2$/GaAs $\delta$-doped samples.
Before annealing, the linewidths remain broad ($\approx$50 GHz) and nearly independent of $d$, indicating that charge- and spin-noise contributions from $V_\textrm{O}$ and Er-Er interactions dominate. 
After annealing, the linewidth decreases with $d$, consistent with prior reports of TiO$_2$ films on silicon,\cite{Singh2024} but two distinct regimes emerge.
Far from the interface, the linewidth decreases by about 8 GHz relative to as-grown films, reflecting the annihilation of $V_\textrm{O}$ centers that promote local charge compensation and broaden linewidth due to excess charge-noise.
Because the magnetic-noise spectral density $S_z(d)$ arising from Ga nuclei is more than six orders of magnitude smaller at this distance, the linewidth is then primarily limited by dipolar Er–Er interactions, yielding $\Gamma_{\textrm{Er-Er}}\approx 42~\si{\giga\hertz}$.
Closer to the interface ($d<20~\si{\nano\meter}$), the linewidth broadens by at least 10~\si{\giga\hertz} after annealing.
This excess broadening indicates increased gallium diffusion into the TiO$_2$ lattice during thermal annealing, which introduces additional electromagnetic noise that outweighs the reduction in $V_\textrm{O}$-related noise.
Er$^{3+}$ ions situated within this interfacial distance also experience a forced electric dipole moment, consistent with a measured excited-state lifetime of 0.2(1)~\si{\milli\second} (see Supplemental Material).
The resulting increased sensitivity of charge- and spin-noise leads to fluorescence quenching that significantly weakens the PLE signal near the interface, even at the high Er concentrations (3000 ppm) employed in our measurements.
The low activation energy for Ga diffusion ($\approx$2.5 eV; diffusion constant $\approx10^{-17}~\si{\centi\meter\squared\per\second}$) previously measured for PLD-grown films facilitates this enhanced intermixing, confirming that interfacial Ga spins dominate the post-anneal dephasing environment.

To model these trends quantitatively, we assume a $V_\textrm{O}$-density gradient, based on reaction-diffusion modeling, that linearly decreases with $d$, i.e., $n_{V_\textrm{O}}(d)\propto d$. 
Whether originating from spin or charge noise, the corresponding contribution of $V_\textrm{O}$ centers to linewidth is $\Gamma_{V_\textrm{O}}(d) = b n_{V_\textrm{O}}^{2/3}(d)$~\cite{Candido2021b, Grant2024}, where $b$ represents the coupling constant for the underlying noise mechanism.
The resulting total linewidth is $\Gamma(d)=\Gamma_{\textrm{Er-Er}}+\Gamma_{V_\textrm{O}}(d)+\Gamma_{\textrm{Ga}}(d)$. 
This analytical form quantitatively reproduces the experimental data in Figure \ref{fig:fig5}(a), showing that $V_\textrm{O}$- and Er–Er-related noise dominate at large $d$, while Ga-spin noise controls the linewidth for $d\lesssim \lambda$.
The larger uncertainties near the interface likely arise from spatially nonuniform Ga diffusion and reduced signal-to-noise ratios due to fluorescence quenching.

Using this distance-resolved linewidth $\Gamma(d)$ together with the strain-induced frequency shift $\omega_\textrm{strain}(d)$, we performed Monte Carlo simulations to model the ensemble spectrum of uniformly-doped A-TiO$_2$/GaAs film shown in Figure~\ref{fig:fig1}(c).
Sampling a million Er$^{3+}$ ions with random spatial positions $d_i$ within the TiO$_2$ layer, the simulations reproduce the lineshape observed experimentally (Figure \ref{fig:fig5}(b)).
Applying the same model to the R-TiO$_2$ grown on GaSb, while adjusting only $\omega_\textrm{strain}(d) = \omega-c/d$, captures the broader, more asymmetric spectrum characteristic of that heterostructure (Figure~\ref{fig:fig5}(c)).
These results demonstrate that a single microscopic framework quantitatively accounts for both strain-driven frequency shifts and noise-limited linewidth broadening across TiO$_2$/III-V interfaces.

Our work establishes a unified microscopic picture that connects interfacial chemistry, strain, and defect dynamics to the optical response of Er$^{3+}$ in TiO$_2$/III–V heterostructures. 
By integrating $\delta$-doped spectroscopy, \textit{ab initio} calculations, noise Hamiltonian modeling, and Monte Carlo simulations, we directly link interfacial structure to ensemble optical behavior. 
The analysis shows that interfacial Ga spins dominate dephasing and linewidth near the oxide/III–V boundary, whereas oxygen-vacancy and Er–Er interactions govern broadening deeper in the film.
The modeling further predicts that the stronger coupling between strain and cation intermixing expected in rutile TiO$_2$ on GaSb results in broader, more asymmetric spectra than those of the anatase TiO$_2$ on GaAs system studied experimentally.
This unified framework quantitatively reproduces linewidth values and both symmetric and asymmetric lineshapes, providing a predictive tool for designing low-decoherence REI/semiconductor interfaces.
Future studies employing two- and three-pulse photon-echo measurements under small external magnetic fields ($B\sim100$ mT) can directly probe the homogeneous linewidth ($T_2$) and spectral-diffusion dynamics as a function of Er-ion distance from the III-V interface.
Such experiments would extend the framework developed here to quantify decoherence processes at the single-ion level and guide the design of scalable hybrid quantum architectures.

\begin{acknowledgement}
R.U. acknowledges funding from the National Science Foundation (NSF) under the CAREER program (ECCS2339469) and a seed grant from the University of Iowa's Office of the Vice President for Research through the P3 Jumpstarting Tomorrow program. 
\text{Ab initio} crystal field calculations, defect formation energies, noise-Hamiltonian modeling, and noise simulations  were supported by the U.S. Department of Energy, Office of Science, Office of Basic Energy Sciences under Award Number DE-SC0023393. 
%M.E.F, D.R.C, H.A.B., and Y.L. acknowledge support from the Department of Energy, Office of Science Basic Energy Science No. DE-SC0023393. D.P. and M.E.F. acknowledge the Center for Energy Efficient Magnonics, an Energy Frontier Research Center funded by the U.S. Department of Energy, Office of Science, under contract DE-AC02-76SF00515.
Target fabrication and pulsed laser deposition were conducted as part of a user project (CNMS2023-B-02072) at the Center for Nanophase Materials Sciences (CNMS), which is a US Department of Energy, Office of Science User Facility at Oak Ridge National Laboratory.
\end{acknowledgement}

\bibliography{refs}

\end{document}